\documentclass[%
superscriptaddress,
 amsmath,amssymb,
prx,
]{revtex4-2}
\usepackage[export]{adjustbox}
\usepackage{graphicx}
\usepackage{color}
\usepackage{xcolor}
\usepackage{ulem}
\usepackage{wasysym}
\usepackage{dcolumn}
\usepackage{bm}
\usepackage{hyperref}
\usepackage{url}
\usepackage[mathscr]{euscript}
\usepackage{mathtools}
\graphicspath{{fig/}}
\definecolor{mygreen}{rgb}{0,0.5,0}
\definecolor{mybrown}{rgb}{0.65,0.16,0.16}

\begin{document}

\title{Dynamics of three-dimensional turbulence from Navier-Stokes equations
}

\author{Katepalli R. Sreenivasan}
\email[]{krs3@nyu.edu}
\affiliation{New York University}
\author{Victor Yakhot}
\email[]{vy@bu.edu}
\affiliation{Boston University}
\affiliation{New York University}

\date{\today}

\begin{abstract}
In statistically homogeneous and isotropic turbulence, the average value of the velocity increment $\delta_r u = u(x+r) - u(x)$, where $x$ and $x + r$ are two positions in the flow and $u$ is the velocity in the direction of the separation distance $r$, is identically zero, and so to characterize the dynamics one often uses the Reynolds number based on $\sqrt{\langle (\delta_r u)^2 \rangle}$, which acts as the coupling constant for scale-to-scale interactions. This description can be generalized by introducing “structure functions” of order $n$, $S_{n} = \langle (\delta_r u)^{n} \rangle$, which allow one to probe velocity increments including rare and extreme events, by considering $\delta_r u{(n)} = O(S_{n}^{1/{n}})$ for large and small $n$. If $S_{n} \propto{} r^{\zeta_{n}}$, the theory for the exponents $\zeta_{n}$ in the entire allowable interval $ -1 < n < \infty$ is one of the long-standing challenges in turbulence (one takes absolute values of $\delta_r u$ for negative $n$), usually attacked by various qualitative cascade models. We accomplish two major tasks. First, we show that the turbulent motion at large scales obeys Gaussian statistics in the interval $0<R_{\lambda}\leq 8.8$, where $R_\lambda$ is the microscale Reynolds number, and that the Gaussian flow breaks down to yield place to anomalous scaling at the universal Reynolds number bounding the inequality above. In the inertial range of turbulence that emerges following the breakdown, the effective Reynolds number based on the turbulent viscosity, $R^*_{\lambda},$ assumes this same constant value of about 9. This scenario works also for the emergence of turbulence from an initially non-turbulent state. Second, we derive expressions for the anomalous scaling exponents of structure functions and moments of spatial derivatives, by analyzing the Navier-Stokes equations in the form developed by Hopf. We present a novel procedure to close the Hopf equation, resulting in expressions for $\zeta_n$ in the entire range of allowable moment-order, $n$, and demonstrate that accounting for the temporal dynamics changes the scaling from normal to anomalous. For large $n$, the theory predicts the saturation of $\zeta_n$ with $n$, leading to two inferences: (a) the smallest length scale $\eta_n = L Re^{-1} \ll L Re^{-3/4}$, where $Re$ is the large-scale Reynolds number, and (b) velocity excursions across even the smallest length scales can sometimes be as large as the large scale velocity itself. Theoretical predictions for each of these aspects are shown to be in quantitative agreement with available experimental and numerical data.
\end{abstract}

\maketitle
PACS numbers 47.27
\vspace{1cm}

\hspace{5cm}{\it The scientific theorist is not to be envied. For Nature, or more precisely},

\hspace{5cm}{\it experiment, is an inexorable  and not very friendly judge of his work. It} 

\hspace{5cm}{\it never says ``Yes" to a theory. In the most favorable cases ``Maybe" and}

\hspace{5cm}{\it in great majority of cases simply ``No".}\\ 

\hspace{10cm} Attributed to A. Einstein \cite{gait}

\section {\bf Introduction}

Turbulent flows are everywhere. Without them life on Earth would be impossible, for they are responsible for the life-sustaining heat and mass transfer. Indeed, it is easy to estimate that, due to absorbed solar radiation, one summer without turbulence would be long enough to boil a four-meter layer of ocean water. Also, it is responsible for the temperature control and ventilation in various modern engineering applications, such as data centers, transportation, chemical technology---in short, for all aspects of our life.

Turbulent flows differ by geometry, smoothness or otherwise of the boundaries, physical noise, details of transitions from laminar to turbulent states, etc. Not surprisingly, they are characterized by transitional Reynolds numbers varying over an enormously wide range. For example, depending on the quality of walls and the quietness of the oncoming stream, the transition Reynolds number in a channel flow can vary over two orders of magnitude. But the excited length scales in turbulence, dependent on the Reynolds number, is very wide at high Reynolds numbers, and so one can ask a logical question: do the small-scale velocity fluctuations at high Reynolds number ``remember'' geometric and other features of their laminar precursors? If they do, different flows must be described by different turbulence theories, involving detailed information on laminar-turbulent transition as well as boundary conditions, which seems to be a mission impossible. If they do not, one can hope for the universality classes and universal theory, qualitatively similar to critical phenomena. Addressing this avenue is the primary purpose of this paper.

To describe fluid motion, Euler \cite{eule} derived his field-equations for ideal (inviscid) fluid which, among other concepts, contained the Bernoulli equation obtained almost twenty years earlier. Later, Navier \cite{navi} and Stokes \cite{stok} came up with the equations for the motion of viscous fluids which, combined with modern numerical and computational methods, have revolutionized modern fluid dynamics. These achievements have solved the problem of all laminar (or regular) flows, including time-dependent ones such as flows past oscillating plates, bluff bodies, etc.
 
Transition to turbulence was experimentally discovered and studied by Reynolds \cite{reyn}, who introduced his famous decomposition of the velocity field into its regular and stochastic components. This work initiated the field of statistical hydrodynamics based on the generally accepted belief that both laminar and turbulent flows can be accurately described by the Navier-Stokes (NS) equations. This factor led in due course to the formulation of various field-theoretical approaches such as Wyld's renormalized expansions in powers of Reynolds number \cite{wyld,moni}, Kraichnan's DIA and LHDIA approximations \cite{krai1,krai2}, Orszag's EDQNM approach \cite{orsz}, etc. They have had considerable successes in the derivation of large-scale turbulence models widely used in engineering computations of flows in complex geometries \cite{laun1,laun2,yakh1,yakh2,bart,yakh3} but have failed in the {\it a priori} prediction of small-scale features dominated by powerful intermittent structures, which occupy increasingly smaller fraction of volume with increasing Reynolds number \cite{sree1} and are responsible for the ever-changing intricate images that have fascinated and inspired humans for hundreds of years.  

The theory of critical phenomena, renormalization group and the $\epsilon$-expansion are among the most remarkable achievements of theoretical physics of twentieth century---perhaps even as a ``philosophical'' paradigm \cite{domb} (see, especially, articles by L.P. Kadanoff and M. Lubin in volume 5, and those by K.G. Wilson and F.J. Wegner in volume 6). The theory is thought to be still incomplete, since it is based on approximations that do not solve the 3D-Ising model exactly \cite{refe}, but the large scale features of critical systems are well-understood in terms of a single divergent length scale $\xi\propto |T-T_{cr}|^{-\nu}$, where $T-T_{cr}$ is the distance from the critical temperature $T_{cr}$ and $\nu$ is a universal exponent. However, the small-scale details of the model are unknown, as the ``scaling hypothesis", based on correlation length $\xi\rightarrow\infty$, is definitely not valid for scales of the order of the mean free path. But what makes this approach successful is that such small scales do not play a significant role at the critical point.    

The problem of hydrodynamic turbulence at large Reynolds numbers is very different, as it appears that all flow scales of the continuum---large, intermediate and small---are dynamically important, and that this dynamics is hard to extract from the Navier-Stokes equations, which are believed to apply to large variety of conditions, including those near the critical point in phase transitions (even though they can be derived directly from microscopic molecular dynamics only for rarefied gases for scales much larger than the mean free path). These equations describe velocity fluctuations generated at the large scale $L$ and dissipated on a scale $\eta$ whose ratio to $L$ becomes vanishingly small as an inverse power of the Reynolds number. No energy is lost in this nonlinear inviscid process of the so-called inertial-range energy transfer which requires strong interactions between fluctuations on different length scales, very different from detailed balancing that is characteristic of microscopic reversibility. The process of formation of small-scale fluctuations out of large scales, often called the ``energy cascade'', introduced as an explicit notion by \cite{onsa} but implied already in Kolmogorov's earlier work \cite{kolm1}. Qualitatively cascades resembles the process leading to the formation of high-energy fractions formed in hadron collisions, where it is responsible for anomalous scaling exponents characterizing the collision process \cite{poly}.

For many years, due to the lack of sufficiently accurate and extensive experimental data, the existence of anomalous scaling in high-Reynolds-number turbulence was set aside from serious consideration. That era is thankfully over, and the existence of anomalous scaling and its possible universality, i.e., independence of scaling exponents on the nature of a flow, has been established as a result of many years of hard work by several groups (see, e.g., Refs.\ \cite{anse,benz,sran,dhru,chen,iyer,iyer1}).

In this paper we consider an infinitely extended fluid stirred (or forced) by a random force at the scale $\Lambda$. We show that as $r/\Lambda \rightarrow \infty$, the large-scale flow that is generated is Gaussian in accordance with the Central Limit Theorem and the dynamic renormalization group worked out by Forster, Nelson and Stephen \cite{fors}. Due to the specific nonlinearity of the NS equations, the small-scale asymptotics $r/\Lambda \rightarrow 0$ corresponds to strong coupling, which makes the turbulence problem very hard. This is one of the reasons for the failure of perturbation expansions in powers of the renormalized Reynolds number. Here, we develop a theory based on the Hopf formulation \cite{hopf} of the NS equations and provide a solution by taking recourse to Polyakov's application of point-splitting for the Burgers equation that is driven by a random force \cite{poly2}, as generalized for the three-dimensional (3D) Navier Stokes equations in \cite{yakh4,yakh5}. In particular, Yakhot \cite{yakh4,yakh4a} supplemented the analysis by an {\it ad hoc} model for pressure-velocity correlations and obtained an explicit expression for anomalous scaling exponents $\zeta_{n}$ for structure functions in the entire range $-1 < n < \infty$: structure functions are moments of velocity increments $\delta_r u = u(x+r) - u(x)$ and the scaling exponents $\zeta_n$ are defined by the expected relation $\langle{(\delta u_r)^n}\rangle \propto r^{\zeta_n}$ in the range $\eta \ll r \ll L$. This model for pressure-velocity correlations \cite{yakh4} was assessed in \cite{kuri} and also in \cite{goto} where the Hopf equation was combined with the Bernoulli equation as an input model for pressure-velocity correlation. The crucial feature here is that we do not use any specific model for pressure-velocity correlations and close the equations via coarse-grained NS equations. We also draw an effective analogy to laminar-turbulent transition and point to supporting evidence from Direct Numerical Simulations (DNS).

The rest of the paper is organized as follows. In Section II, we show using the Fourier space representation that the NS equations driven by a random force at the wavenumber $\kappa=\frac{2\pi}{\Lambda}$ obeys Gaussian statistics in the limit of the wavenumber $k/ \kappa \rightarrow 0$, in accordance with \cite{fors}. This Gaussian state corresponds to two physically different situations: (a) the large scale of fully turbulent flows, and (b) the weakly fluctuating state prior to transition to the turbulent state. In both cases the Gaussian state yields place abruptly to anomalous scaling. Section III shows that in the inviscid inertial range, $\eta \ll r \ll \Lambda$, there is a scale-independent effective Reynolds number $R^*_{\Lambda}=8.8$ at which transition from the Gaussian state to the inertial range takes place. In Section IV, the Hopf equations are derived and the point-splitting procedure is presented as a step towards ``opening" the Hopf equation. Sections V evaluates the coarse-grained NS equations for pressure contributions from the coarse-grained Navier-Stokes equations and obtains scaling exponents to moments of all orders. We show in Section VI that ignoring the time-derivative in the NS equations leads to the Kolmogorov scaling $\zeta_{2n}=2n/3$, and that one obtains anomalous scaling when both spatial and temporal dependencies are included. In Section VII, The predictions of the present theory are substantiated by comparison with various experiments and simulations for both (a) and (b) above. The final Section VIII is devoted to a brief recapitulation of results and conclusions.

\section{\bf The model. Large-scale Gaussian Statistics}  

Flow of Newtonian fluids can be described by the NS equations subject to boundary and initial conditions (with the density $\rho$ set to unity without loss of generality),  
\begin{equation}
\partial_{t}{\bf u}+{\bf u\cdot\nabla u}=-\nabla p +\nu\nabla^{2}{\bf u} + {\bf f}
\end{equation}
and the incompressibility condition given by $\nabla\cdot {\bf u}=0$. The random white-in-time Gaussian forcing ${\bf f}$ is defined by the correlation function \cite{fors,yakh1}
\begin{equation}
\langle{f_{i}({\bf  k},\omega)f_{j}({\bf k'},\omega')\rangle}= (2\pi)^{d+1}D_{0}(k)P_{ij}({\bf k})\delta({\hat{k}+\hat{k}'}),
\end{equation}
where the four-vector $\hat{k}=({\bf k},\omega)$ and projection operator is given by $P_{ij}({\bf k}) = \delta_{ij}-\frac{k_{i}k_{j}}{k^{2}}$. Here we are interested in the case $D_{0}(k)\neq 0$ only in the interval close to $k\approx \Lambda \approx 2\pi/L$. Taking the Fourier transform of (1) as
\begin{widetext}
\begin{equation}
u_{l}({\bf k},\omega)=G^{0}f_{l}({\bf k,\omega})-
\frac{i}{2}G^{0}{\cal P}_{lmn}\int u_{m}({\bf q},\Omega)u_{n}({\bf k-q},\omega-\Omega)d{\bf Q}d\Omega,
\end{equation}
\end{widetext}
where $G^{0}=(-i\omega+\nu k^{2})^{-1}$, ${\cal P}_{lmn}({\bf k})=k_{n}P_{lm}({\bf k})+k_{m}P_{ln}({\bf k})$, we introduce the zeroth-order solution ${\bf u}_{0}=G^{0}{\bf f} \propto k\sqrt{D_{0}}$, so that ${\bf u}=G^{0}{\bf f}+{\bf v}$, we derive the equation for the perturbation velocity ${\bf v}$ to be
\begin{widetext}
\begin{eqnarray}
v_{l}(\hat{k})=-\frac{i}{2}G^{0}(\hat{k}){\cal P}_{lmn}({\bf k})\int v_{m}(\hat{q})v_{n}(\hat{k}-\hat{q})d\hat{q}\nonumber\\
-\frac{i}{2}G^{0}(\hat{k}){\cal P}_{lmn}({\bf k})\int  [v_{m}(\hat{q})G^{0}(\hat{k}-\hat{q})f_{n}(\hat{k}-\hat{q})+ G^{0}(\hat{q})f_{m}(\hat{q})v_{n}(\hat{k}-\hat{q})]d\hat{q} \nonumber \\
-\frac{i}{2}G^{0}(\hat{k}){\cal P}_{lmn}({\bf k})\int G^{0}(\hat{q})f_{m}(\hat{q})G^{0}(\hat{k}-\hat{q})f_{n}(\hat{k}-\hat{q})d\hat{q}.
\end{eqnarray}
\end{widetext}
 
Because of the nonlinearity of the NS equations, the modes from the large-scale range $k\rightarrow 0$, where the forcing ${\bf f}(k)=0$, are populated by the ``effective forcing'' ${\bf F}$ \cite{fors,yakh1} as
\begin{equation}
\langle{F_{i}(\bf k)F_{j}({\bf k'})\rangle}\approx 2D_{L}k^{2}P_{ij}({\bf k})\delta({\bf k+k'})+O(k^{4})
\end{equation}
where (see \cite{lesl})
\begin{equation}
D_{L}=(2\pi)^{d+1}D_{0}\frac{0.155}{\Lambda^{5}},
\end{equation}
which is small in the limit $\frac{k}{\kappa}\sqrt{\frac{D_{0}}{D_{L}}}\rightarrow 0$. In this scale range of weak coupling $|F(k)|^{2}\propto k^{2}D_{0}\rightarrow 0$, the low-order perturbation expansion is accurate, and the flow is in ``thermodynamic equilibrium", stabilized by a small {\it induced} viscosity $\nu_{eff}=\nu_{0}+\delta\nu$ where $\nu_0$ is the bare viscosity and $\delta \nu\propto D_{L}$. In an infinitely extended fluid satisfying the limiting condition $k^{2}D_{0}\rightarrow 0$, the flow that is generated is Gaussian by virtue of the Central Limit Theorem and dynamic renormalization group \cite{fors,yakh1,lesl}, independent of the statistics of the force $\bf f$ introduced earlier.

To conclude this section devoted to the large-scale behavior at $k \ll \Lambda$, we note that in the strong turbulence range $k\gg \Lambda$, the induced forcing evaluated in the low-order approximation is $\langle F^{2} \rangle \propto k^{-3}$. This model was introduced in \cite{dedo}. It is clear that the entire turbulence production in the force-driven turbulence has a logarithmic divergence, compensated by the infrared range and higher nonlinearities that are not directly relevant in our considerations.

\section{\bf  The model. Small-scale asymptotics and the ``critical Reynolds number"}  
In the large wavenumber regime often referred to as the ultra-violet regime, $k\gg \kappa$ with $k^{2}D_{0}\rightarrow\infty$, the dynamics are far from equilibrium and perturbation expansions break down. In this range, the flow is driven by an effective forcing
and the energy is dissipated by viscosity in the so-called ``dissipation range". While the induced field is weak in the weak-coupling range $k\rightarrow 0$, it is strong in the limit $\frac{k}{\kappa}\rightarrow \infty$ and is responsible for various nonlinear effects considered below. One can attempt to extend Wyld's expansion into this range but the task is difficult because such expansions are haunted by ``infrared" divergences \cite{krai1,krai2,fors}.

In summary, at large scales $k \ll \kappa$, the velocity fluctuations in the flows driven by the forcing centered at $\kappa =\frac{2\pi}{L}\equiv \Lambda$ are Gaussian, justifying various one-loop approximations to renormalized perturbation expansions. The modes with $k>\kappa$ is populated by strong nonlinearity, and the proper matching between viscous and inertial range effects becomes a dominant mechanism responsible for anomalous scaling and strong departures from Gaussian statistics.
 
We stress the lack of small parameter in the inertial range by the following argument \cite{land}. For the velocity increment $u_r$ in the inertial range where dissipation by the molecular viscosity is negligible, the effective or turbulent viscosity governing the dynamics of large scales ($\Lambda \gg l\gg r\gg \eta$) is given by $\nu_{T}\approx r u_{r}$. In this case, the dimensionless coupling constant is the Reynolds number  
\begin{equation}
Re(u_r)=\frac{r u_r}{\nu_{T}}=const=O(1),
\end{equation}
which is not small anywhere in the inertial range $\eta\leq r\leq L$. In the same spirit\footnote{The equation for the $x$-component of the nonlinearity in the Navier-Stokes equation (1) is equal to
\begin{equation}
\partial_{t}v_{x}+\frac{1}{2}v_{x}^{2}+v_{y}\partial _{y}v_{x}+v_{z}\partial_{z}v_{x}\approx \partial_{t}v_{x}^{>}+\frac{1}{2}v^{>}_{x}^{2}+v^{>}_{y}\partial _{y}v^{>}_{x}+v^{>}_{z}\partial_{z}v^{>}_{x} +\partial_{r}\nu_ {T}(r)\partial_{r}v^{>}_{x}+O(v^{{2}}).
\end{equation}
By Galilean invariance, small-scale elimination does not modify coefficients in the convection terms in the equation  
$v_{r}$ but renormalizes the dissipation contribution by introducing $\nu_{T}(r)$. The physics involves the tendency to generate longitudinal sharp steps. The x-y and x-z terms prevent the shock formation by excitation of transverse $y$ and $z$ components of velocity field. This manifests itself in the appearance of effective viscosity $\nu_{T}(r)$.}, one derives the expression for turbulent viscosity as
\begin{equation}
\nu_{T}\equiv \nu(\Lambda)\approx 0.084\frac{K^{2}}{\varepsilon}; \hspace{2cm}10 \nu^{2}(\Lambda)\times \Lambda^{2}=K^2,
\end{equation}
where $K \approx u_r^2/2$ is kinetic energy of the turbulent motion on the inertial scales $r\leq L$, and $\varepsilon$ is the mean rate of kinetic energy dissipation. Landau and Lifshitz \cite{land} show that the turbulent viscosity is $\nu_{T}(r)\propto r \langle {u_r^2} \rangle ^{1/2}$ given by the expression above.

An important point needs to be made. The relations (8) have been derived by the small-scale elimination from the interval $\eta <r< \Lambda$ and, therefore, give effective viscosity acting on large-scale fluctuations $k<\frac{2\pi}{r}$. Further, the effective Reynolds number of single-scale to multi-scaling transition, based on the Taylor scale and turbulent parameters
\begin{equation}
R_{\lambda,r}=2K\sqrt{\frac{5}{3{\cal E}\nu(\Lambda)}}= 2\sqrt{\frac{5}{3\times 0.084}}   \approx 8.8
\end{equation}
is scale-independent in the inertial range and is very close to the Reynolds number of transition from the Gaussian to anomalous regime of the velocity field.

This simple relation (9), combined with the definition of turbulent viscosity $\nu_{T}\propto \frac{\langle (\delta_{r}u)^{4} \rangle}{{\cal E}}$, valid in homogeneously or smoothly excited flows, gives
\begin{equation}
\langle (\delta_{r}u)^3 \rangle \propto {\cal E}r,
\end{equation}
which resembles the celebrated Kolmogorov relation for velocity increments in the inertial range of homogeneous and isotropic turbulence. Thus, unlike in the theory of critical phenomena, the constant energy flux from large to  small scales in turbulence makes all scales dynamically relevant, which leads to a gross violation of ``scaling hypothesis''. This is the main reason for the failure of field-theoretical methods like Wyld's expansion \cite{wyld}, Kraichnan's DIA and LHDIA \cite{krai1,krai2} and others (see, for example, \cite{moni}).

It appears from this discussion that the infrared limit can be understood by the perturbation theory, so our goal in the next section is to develop the theory in the strong coupling ultraviolet range, where $k>\kappa$ and $k^{2} D_{0} \rightarrow \infty$.
 
\section{IV. Hopf equation, point splitting and the scaling exponents}

\subsection{The Hopf equation}
As was shown by Kolmogorov \cite{kolm2} in the high-Reynolds-number limit, the third-order structure function $S_{3}(r) = -\frac{4}{5} \varepsilon r$ in the inertial range of isotropic and homogeneous turbulence. This is the celebrated $4/5$-ths law. Our goal is to discuss all even moments $S_{2n}=\langle{(u(x)-u(x+r))^{2n}\rangle}$ and their scaling exponents $\zeta_{2n}$ where $S_{2n} \propto r^{\zeta_{2n}}$. As already mentioned a large body of experimental work \cite{sran} has demonstrated that $\zeta_{2n}$ are  different from Kolmogorov's theory, $S_2n(r) \propto r^{2n\zeta}$, where the ``local'' velocity increment $[S_{2n}(r)]^{(1/2n)} \propto r^{\zeta}$ \cite{kolm1}. This is the ``normal scaling". The problem of determining $\zeta_{2n}$, if they are different from $2n\zeta$, is important because they determine the behavior of large excursions or the tails of the distribution of $\delta_r u$.

We address the problem of determining $\zeta_{2n}$ using Hopf equation \cite{hopf}, adopting the approach used by Polyakov \cite{poly2} for the description of Burgers equation driven by a random force, later generalized in Refs.\ \cite{yakh2,yakh3} to the case of 3D Navier-Stokes turbulence in the evaluation of velocity increments $\bf {U}\equiv \delta_{r}u={\bf u(x+r)-u(x)}$; see also Ref.\ \cite{goto}. We will show that, due to large nonlinearity, the probability of large-amplitude and rare fluctuations is much larger than that for the Gaussian determined by the second moment. It has been known for some time that the volume fraction occupied by these very local events is small and thus the total expectation value of these events is small \cite{sree1}.
 
The origins of the Hopf equation for the generating function $Z=<e^{\lambda\cdot U)}>$ goes back to Ref.\ \cite{hopf}. The relevant equation becomes
\begin{equation}
\frac{\partial Z}{\partial t}+\frac{\partial^{2}Z}{\partial \lambda_{\mu}\partial r_{\mu}}=I_{f}+I_{p}+D
\end{equation}
\begin{equation}
I_{p} =-\lambda\cdot \langle e^{\lambda\cdot u_{r}}[\nabla_{2}p(x_{2})-\nabla_{1}p(x_{1})] \rangle
\end{equation}
\begin{equation}
 D=\nu_{0}\lambda\cdot \langle [ \nabla_{2}^{2}v(x_{2})-\nabla_{1}^{2}v(x_{1})]e^{\lambda\cdot u_{r}}\rangle  
\end{equation}
\begin{equation}
 I_{f}= \langle\lambda\cdot {\bf \nabla f}e^{\lambda\cdot u_{r}}\rangle.
\end{equation}

The main difficulty in solving this equation is that it is not closed because of the dissipation and pressure terms (for a more detailed discussion see Monin and Yaglom \cite{moni}) and cannot be solved unless additional  information can be extracted from the NS equations. In the inertial range where  $\nu_0\rightarrow 0$ and $r=|x_{1}-x_{2}|\gg \eta$, the dissipation contribution $D=0$.  This is not so in the theoretically harder limit $r\approx \eta$ or $x_{1}\rightarrow x_{2}$, as discussed below.

In the rest of the paper we define $S_{m,n} = \langle (\delta_r u)^m (\delta_r v)^n \rangle$, where $u$ and $v$ are velocity fluctuations in the direction of the separation and perpendicular to it, respectively. By symmetry, $v$ is the same in any direction perpendicular to $r$. The equations for the even-order moments \cite{yakh1,yakh2,yakh3} are
\begin{eqnarray} \frac{\partial S_{2n,0}}{\partial r}+\frac{d-1}{r}S_{2n,0}-\frac{(d-1)(2n-1)}{r}S_{2n-2,2}+ (2n-1) \langle \delta_{r} (\partial_{x}p)\times (\delta_{r}u)^{2n-2} \rangle=\\ \nonumber
{\cal P}[1-\cos(\frac{r}{L})]\alpha_{n,d}S_{2n-3,0}+(2n-1) \langle \delta_{r} \hat{a}(\delta_{r}u)^{2n-2} \rangle
\end{eqnarray}
where $\alpha_{n,d}=2(2n-1)(2n-2)/d$, $d$ being the dimensionality of the fluid system. By energy conservation, we have ${\cal P}={\cal E}$ and $\hat{a}$ is the Lagrangian acceleration of the fluid particle.

In particular, the last term in the above equation with $\delta_{r}\hat{a}=\hat{a}(x+r)-\hat{a}(x)$ with the increment over $r$ in the inertial range, is unknown and has to be treated separately. In the vicinity of dissipation scale $\eta$, the characteristic time is $\nu/(\delta_{\eta}u)^{2}$ and so we have
\begin{equation}
\delta_{\eta}\hat{a}=\frac{(\delta_{\eta}u)^{3}}{\nu}
\end{equation}
and
\begin{equation}
\langle \delta_{r}\hat{a}(\delta_{r}u)^{2n-2} \rangle \approx \frac{1}{\nu} \langle (\delta_{\eta}u)^{3}(\delta_{r}u)^{2n-2} \rangle.
\end{equation}

First we consider the inertial range, for which $\nu\rightarrow 0$, $r\gg \eta$, and $\frac{r}{L}\rightarrow 0$, implying that the dissipation contribution given by $\nu \langle \delta_{r}\hat{a}(\delta_{r}u)^{2n-2} \rangle=0$. Thus,
\begin{equation}
\frac{\partial S_{2n,0}}{\partial r}+\frac{d-1}{r}S_{2n,0}-\frac{(d-1)(2n-1)}{r}S_{2n-2,2}=-(2n-1) \langle \delta_{r} (\partial_{x}p)\times (\delta_{r}u)^{2n-2}\rangle
\end{equation}
and
\begin{equation}
-\frac{\partial p}{\partial x} =\frac{\partial v_{x}}{\partial t}+\frac{1}{2}\frac{\partial v_{x}^{2}}{\partial x}+v_{y}\frac{\partial v_{x}}{\partial y} +v_{z}\frac{\partial v_{x}}{\partial z}.
\end{equation}

Substituting (19) in (18), for $2n=2$ and $2n=3$, the pressure contribution in the incompressible case disappears and one obtains the K41 theory for $S_{3,0}(r)$ and $S_{1,2}$ as
\begin{equation}
\frac{\partial S_{3,0}}{\partial r}+\frac{d-1}{r}S_{3,0}-2\frac{d-1}{r}S_{1,2}=-(1)^{d}\frac{4}{d}{\cal P},
\end{equation}
\begin{equation}
S_{3,0}=(-1)^{d} \frac{12}{d(d+2)}{\cal P} r
\end{equation}
and  
\begin{equation}
\frac{1}{r^{d+1}}\frac{d}{dr}r^{d+1}S_{1,2}=(-1)^{d}\frac{4}{d}{\cal P}.
\end{equation}
These relations are exact in the limit $\nu\rightarrow 0$ and yield the well-known results (for $d=3$)  
\begin{equation}
\zeta_{2,0} = \zeta_{0,2},~~~~~ \zeta_{3,0} = \zeta_{1,2},
\end{equation}
in addition to Kolmogorov's celebrated $4/5$-ths law and the relation $S_{3,0}/S_{1,2} = 3$.

\subsection{Viscous effects, point-splitting, and matching condition}

Equations (11-15), containing both pressure and dissipation contributions, are not closed for general $n$. As we have shown, the dissipation contribution is small in the inertial range $r\gg\eta$ and can be neglected. However, it remains important in the range $r\approx \eta$ and must be accounted for. This is done by asymptotic matching of structure functions from inertial and dissipation intervals, leading to important dynamic consequences, as studied below (see also \cite{yakh3}).
 
As discussed in Landau and Lifshitz \cite{land}, if $\delta_{\eta}u\approx u(x+\eta)-u(x)$, the dissipation scale $\eta$ is defined by the condition $Re_{\eta}=O(1)=\eta (\delta_{\eta}u )/\nu$, so that $\eta= O(\nu/(\delta_{\eta}u))$. Thus, the simple algebra, based on expressing spatial derivatives in terms of point-splitting as above, gives
\begin{equation}
\partial_{x} {u}\approx (\delta_{\eta}u)^{2}/\nu.
\end{equation}
Therefore
\begin{equation}
(\frac{L}{u_{0}})^{2n}\langle(\frac{\partial v_{x}}{\partial x})^{2n} \rangle=Re^{2n}(\frac{\eta_{4n}}{L})^{\zeta_{4n}} \equiv Re^{\rho_{2n}},
\end{equation}
where $Re=u_{0}L/\nu$, where $u_0$ is a large scale property such as its root-mean-square value. We then get
\begin{equation}
\frac{L^{n}}{u_{0}^{3n}}{\langle\cal E}^n\rangle=Re^{n}
\frac{\langle(\delta_{\eta}u)^{4n}\rangle}{{u_{0}}^{4n}}
=Re^{n}S_{4n}(\eta_{4n})=Re^{n}(\frac{\eta_{4n}}{L})^{\zeta_{4n}}  \equiv Re^{d_{n}},
\end{equation}
where the scaling exponents $d_{n}$ and $\rho_{n}$ are yet to be derived from an {\it a priori} theory. From the application of Eq. (24) which defines the point-splitting procedure (see \cite{yakh5}), we have
\begin{equation}
\frac{S_{2n}(\eta_{2n})}{\eta_{2n}}\approx \frac{S_{2n+1}(\eta_{2n})}{\nu},
\end{equation}
leading to
\begin{equation}
\frac{\eta_{n}}{L}=Re^{\frac{1}{\zeta_{n}-\zeta_{n+1}-1}}.
\end{equation}

For the case of ``normal scaling"  with $\zeta_{n}=n\zeta$, this expression gives Kolmogorov-like $n$-independent relation
\begin{equation}
\frac{\eta_{n}}{L}=Re^{\frac{-1}{\zeta+1}}.
\end{equation}
By requiring (28) to coincide with $\frac{\eta_{2}}{L}=Re^{-3/4}$, we get the Kolmogorov result, $\zeta=1/3$. We also have two important additional relations
\begin{equation}
\rho_{n}=n+\frac{\zeta_{2n}}{\zeta_{2n}-\zeta_{2n+1}-1},
\end{equation}
and
\begin{equation}
d_{n}=n+\frac{\zeta_{4n}}{\zeta_{4n}-\zeta_{4n+1}-1}=\rho_{2n}-n,
\end{equation}
where $\rho_n$ and $d_n$ are defined by Eqs.\ (25) and (26), respectively. Note that, the so-called dissipative anomaly \cite{sree84} corresponds to $d_1 = 0$, which yields the result $\zeta_5 = 2\zeta_4 -1$; from the point of view of the present theory, the latter result may thus be regarded just as basic as dissipative anomaly. 

There are three sets of unknowns obeying two equations and so the above relations for exponents of derivative moments and dissipation rate are not closed. They will be closed and the equations derived in this section earlier will be solved below.

\section{Scaling exponents to all orders}
The quantitatively reasonable success of the simple statement $S_{2}(r)=(S_{3}(r))^{\frac{2}{3}}\propto r^{\frac{2}{3}}$ inspired many followers, both theorists and experimentalists, to assume that
\begin{equation}S_{2n,0}(r) \equiv \langle {(u(x+r)-u(x))^{2n}} \rangle=A_{2n,0}r^{\zeta_{2}n}=A_{2n,0}r^{\frac{2n}{3}},
\end{equation}
where $A_{2n,0}$ could depend on large scale properties.  The closeness of the exponent to 2/3 introduced Kolmogorov's energy spectrum $E(k)\propto {\cal P}^{\frac{2}{3}}k^{-\frac{5}{3}}$, dominating turbulence theory for many years. Data in hydrodynamic turbulence, plasma turbulence and even in certain cosmological situations confirmed the approximate validity of this spectral form but considerable work---for example, Ref.\ \cite{iyer1}---has shown that there are (modest) corrections to the 5/3 exponent. Our goal is to derive not only the second order exponent but exponents to all orders.  

For that purpose, we use methods developed earlier \cite{sina,yakh6,yakh7} for the problem of passive scalar advected by a random velocity field, generalized to the calculation of single-point probability density functions (PDFs) and moments of vorticity and dissipation rates in turbulence. Relatively recently, a similar approach, combined with qualitative input from coherent structures, has been developed for the moments of velocity increments leading to quantitative derivation of anomalous scaling exponents \cite{yakh1,goto}. Here, we improve on the past methods and derive an assumption-free quantitative expression for the exponents of structure functions to all orders, as well as for the moments of the dissipation rate in strong turbulence. We derive results explicitly for even orders but make comparisons also with negative moments for absolute values of velocity increments; the same principle should also apply to odd moments.

To derive structure functions $S_{n}(r)$ for $n\neq 3$, we need an expression for pressure-velocity correlation function in the limit $\nu\rightarrow 0$. The two terms on the left side of Eq.\ (18) are $O((\delta_{r}u) ^{2n}/r)$ and $O((\delta_{r}u))^{2n-2}(\delta_{r}v^{2})/r$ and the scaling exponents of structure functions are determined by the coefficients in (25)-(26), not by dimensional considerations. This places very strict constraints on the the pressure-velocity correlation in (18), whose non-locality is the source of much complexity. We observe that the pressure contribution of (18) can only modify coefficients in the left side, but it is clear that, all powers other than $2$ breaks the balance in both small and large-$n$ limits.  

To close Hopf equation, we have to calculate (see Eq.\ (18))
\begin{equation}
\delta_{r}\frac{\partial p(x)}{\partial x}=\frac{\partial p(x+r)}{\partial( x+r)}-\frac{\partial p(x)}{\partial x}.
\end{equation}
where $\delta_{r}p=\delta_{r}[\frac{\partial_{i}\partial _{j}}{\partial^{2}}u_{i}u_{j}]$. Since  $r \gg \eta$, this relation involves two vastly different length scales and is hard to deal with. Still, based on qualitative reasoning, coming from the Navier-Stokes equations, we have
\begin{equation}
\delta_{r}\nabla p=\delta_{r}[\nabla \frac{\nabla_{I}\nabla_{j}u_{I}u_{j}}{\nabla^{2}}]\approx
\frac{1}{2}\frac{\partial
(\delta_{r}u)^{2}}{\partial r}+O(v^{2}).
\end{equation}

Now we present a dynamical derivation of this relation. First, in the vicinity of transition $R_{\lambda}\approx 8.8$, a slowly varying large-scale velocity field ${\bf u}_{0}(X)$ is formed (see \cite{yakh8} and Section II). With further increase of the Reynolds number, the growing large-scale fluctuations $\bf u_0$ are stabilized by effective viscosity (8), originating from the relatively high-frequency velocity fluctuations $\bf v$, making the large scale fluctuations quasi-steady. (In the literature on turbulence decay, this concept is often called ``permanence of large eddies".) Therefore, the equation for turbulent fluctuations in ${\bf u}={\bf u_{0}}+{\bf v}$ with ${\bf v}({\bf k})$ and $k > \Lambda$, derived in \cite{yakh8}, is
\begin{equation}
\partial_{t}{\bf v}+{\bf v}\cdot\nabla {\bf v}=-\nabla p+{\bf f}+\nu_{0}\nabla^{2}{\bf v},
\end{equation}
where
\begin{equation}
{\bf f}={\bf  f}_{1}+{\bf f}_{2}+\bf{f}_{3}=-{\bf u}_{0}\cdot\nabla {\bf v}-{\bf v}\cdot \nabla {\bf u_{0}}+(\nu_{0}-\nu_{tr})\nabla^{2}{\bf u}_{0}.
\end{equation}
In the local frame of reference moving with the slowly varying velocity ${\bf u}_{0}$, we have
\begin{equation}
-{\bf \delta p_{x}(X,r)=\delta u^{>}_{t}  + \delta u^{>}(X,r)\partial_{r}\delta u^{>}(X,r)=\delta u^{>}_{t}+\frac{1}{2}\frac{\partial}{\partial r}(\delta u^{>})^{2},}
\end{equation}
with $\delta\equiv \delta_{r}$ and 
\begin{equation}
<\delta_{r}\partial_{x}p(x)|U^{2n-2}>=\int_{-\infty}^{\infty}<\delta u_{t}+\delta_{r}\partial_{x}p(x)|U,r>U^{2n-2}P(U,r)dU>,
\end{equation}
and $u^{>}$ is the velocity field defined on the interval $[\eta,r]$. Substituting (38) for pressure gradient into the equation (18) closes the Hopf equations. It is in this way that we have removed the infrared fluctuations from the problem. Consequently, in the moving frame of reference, the large scale, slow component disappears from the homogeneous equation of motion which becomes the equation for a flow in the inertial range of scales, $r\geq \eta$. In the early turbulence literature (e.g., see Kadomtsev \cite{kado}), this approach, dealing  with the infrared divergences, was called ``random Galilean transformation''.  

To study the dynamics on the scales of interest $|x_{1}-x_{2}|\approx r$, we have to eliminate, in addition, the smallest scale fluctuations from the interval $[\eta,r].$ This is achieved by using the ``smoothing procedure" based on ``turbulent viscosity'' (8) which is the result of small-scale elimination (or coarse-graining) procedure from the interval $\eta \leq x \leq r$. In other words, it describes the effective viscosity acting on entities averaged over high-frequency fluctuations in the interval $\eta < x \leq r$. Therefore, in the effective (coarse-grained) Navier-Stokes equations, the operator $\delta_{r}=\hat{x}$ where $\hat{x}\approx \delta_r$ stands for the coarse-grained local coordinate. As a result,
\begin{equation}
\frac{\partial u_{r}}{\partial t}+\frac{1}{2} \frac{{\partial (u_{r})^{2}}}{\partial r} +O(v^{2})
=\frac{\partial}{\partial r}\nu_{T}(r)\frac{\partial}{\partial r}  u_{r} -\frac{\partial p_{r}}{\partial r}
\end{equation}
where the turbulent viscosity $\nu_{T}(r) \approx r u_{r} \approx 0.085 \frac{K^{2}}{\epsilon} \approx 0.025(\delta_{r}u)^{4}/\epsilon$, and $\delta_{r}u\equiv u_{r}$. The procedure leading to this equation, though cumbersome, is described in detail in Refs.\ [12], [14] and [35]. 
 
As already mentioned, the basic idea of removing the infrared divergences using Lagrangian description with random Galilean transformation has been proposed for many years. But the major difference between those early efforts and present is that the equations for the moments, with which we are dealing here, contain an infinite series of Wyld's diagrams coming from all orders of renormalized perturbation expansion, describing interactions of all orders. This feature is crucial, for, indeed, while  large-scale fluctuations do ``carry'' small scales, they account only for a part of the dynamics involving coherent structures, and the origins of anomalous exponents cannot be unearthed from one-loop diagrams, or even an infinite set of them. 

Thus, after removing fluctuations on length-scales $l>r$ and $l<r$, we are left with the following equation, defined on an interval $l\approx r$ for $p_{r}=p(r)-p(0)$ and $u_{r}=u(r)-u(0)$, averaged over positions in the flow:
 \begin{equation}
\frac{\partial u_{r}}{\partial t}+\frac{1}{2}\frac{\partial u_{r}^{2}}{\partial r} +O(v^{2})= \frac{\partial}{\partial r}\nu_{T}(r)\frac{\partial}{\partial r} u _{r} - \frac{\partial p_{r}}{\partial r}.
\end{equation}
Here the $O(v^{2})$ contribution includes $O(u_{y}\partial_{y}u_{x})$. Substituting (38) into (18) gives a closed equations for structure functions as 
\begin{equation}
\frac{\partial S_{2n,0}}{\partial r}+\frac{d-1}{r}S_{2n,0}-\frac{(d-1)(2n-1)}{r}S_{2n-2,2}=-(2n-1)<
-\frac{\partial u_{r}}{\partial t}-\frac{1}{2}\partial_{r} u_{r}^{2}+\partial_{r}\nu_{T}(r)\partial_{r}u_{r}+O(v^{2})]\times u_{r}^{2n-2}>
\end{equation}

Since $\partial_{t}u_{r}\propto \partial_{r}u_{r}^{2}$ in the inertial range, we define $a= 
-\frac{\partial u_{r}}{\partial t}-\frac{1}{2}\partial_{r} u_{r}^{2}$ and take into account
\begin{equation}
a\frac{\partial (\delta u)^{2}}{\partial r}(\delta u)^{2n-2}=\frac{a}{n}\frac{\partial S_{2n}}{\partial r} \hspace{1cm} {\rm and}  \hspace{1cm} b=<{(\delta v}^{2})( u_{r})^{2n-2}>\equiv S_{2n-2,2},
\end{equation}
as well as the relation based on (8), namely $\partial_{r}\nu_{T}(r)\partial_{r}u_{r} \approx 0.025\partial_{r}(u_{r}^{2})$. From  (41)-(42) we finally obtain
\begin{equation}
\zeta_{2n,0}=\frac{(2n-1)(2-b)\frac{A_{2n-2,2}}{A_{2n,0}}-2}{(1-2a)n+a}n.
\end{equation}
This relation is subject to the constraint $a-b=\frac{1}{4}$ to guarantee that $\zeta_{3}=1$. Momentarily neglecting the time-derivative and turbulent viscosity in (41)-(43) gives $a=\frac{1}{2}$ and $b=\frac{1}{4}$. With this information, we are now prepared to analyze the closed equation (41)-(42).  

\section{The role of temporal derivative in the Navier-Stokes Equations}  

\subsection{Steady state:  $\frac{\partial u_{r}}{\partial t}=0$. The Kolmogorov or normal scaling} 

The equation for the $x$-component of nonlinearity in  the Naver-Stokes equation (1) is equal to
$$\partial_{t}v_{x}+\frac{1}{2}v_{x}^{2}+v_{y}\partial _{y}v_{x}+
v_{z}\partial_{z}v_{x}\approx \partial_{t}v_{x}^{>}+\frac{1}{2}\partial_{x}(v_{x}^{2})+v^{>}_{y}\partial _{y}v^{>}_{x}+v^{>}_{z}\partial_{z}v^{>}_{x} +\partial_{r}\nu_ {T}(r)\partial_{r}v^{>}_{x}+O(v^{{2}})$$
By translational invariance small-scale elimination does not modify coefficients in front of convection terms 
but renormalizes dissipation contribution by introducing $\nu_{T}(r)$ leading to  equation (41) above.  This justifies the coefficient $a=\frac{1}{2}$  in (43).

The physics in (41)-(43) describes the tendency to generate longitudinal sharp steps. The x-y and x-z terms prevent  the shock formation by excitation of transverse $y$ and $z$ components of velocity field.  This is manifested by the appearance of effective viscosity  $\nu_{T}(r)$ in equation (41). From (41)-(43) we readily obtain
\begin{equation}
\zeta_{2n,0}=\frac{(2n-1)(2-b)\frac{A_{2n-2,2}}{A_{2n,0}}-2}{(1-2a)n+a}n.
\end{equation}
This relation is subject to the constraint $a-b=\frac{1}{4}$ to guarantee that $\zeta_{3}=1$. Neglecting  for a time being the time -derivative and turbulent viscosity in (41)-(43) we obtain  $a=\frac{1}{2}$ and $b=\frac{1}{4}$ . Equation (44) simplifies to 
\begin{equation}
\zeta_{2n,0}=[(2-b)\times (2n-1)\frac{A_{2n-2,2}}{A_{2n,0}}-2]\times 2n.
\end{equation}
For Gaussian or ``normal'' (i.e., linear) scaling,  $(2n-1)\frac{A_{2n-2,2}}{A_{2n,0}}=\zeta_{0}$= constant, so extrapolating the low-order relation $(2n-1)\frac{A_{2n-2,2}}{A_{2n,0}}=\frac{4}{3}$ into the large -${n}$ interval, one obtains Kolmogorov's exponents $$\zeta_{2n}=2n/3; \hspace{2cm}  \zeta_{3}=1.$$ 

In the past Kolmogorov's relations were mostly obtained from dimensional arguments, and the associated physics was contained in the so-called cascade models qualitatively describing energy cascade from the large scale through the inertial range to the small ones where it is dissipated into heat. No such process has been assumed in the above derivation.

\subsection{B. Anomalous scaling and saturation of exponents}  
Keeping the time derivative and turbulent viscosity 
$$\partial_{r}\nu_{T}(r)\partial_{r}u_{r}\approx 0.025\partial_{r} u_{r}^{2}$$ 
shifts numerical coefficients in Eq.\ (41), evaluated above,  to $a=\frac{1}{2}-0.025=0.475$, which yields $b=0.225$ from the constraint just below Eq.\ (43).  The exponents $\zeta_{2n,0}$ are modified to
\begin{equation}
\zeta_{2n,0}= \frac{(\frac{9}{4}-a)\times (2n-1)\frac{A_{2n-2,2}}{A_{2n,0}}-2 }{(1-2a)n+a}\times n.
\end{equation}

The expression $(2n-1)A_{2n-2,2}/A_{2n,0} = 4/3$ for low moments; in fact it is an approximate constant in the entire interval tested in numerical simulations. In the limit $n\rightarrow{\infty}$, we are thus led to
\begin{equation}
\zeta_{2n}\rightarrow 7.3.
\end{equation}

We point out that at low enough Reynolds numbers where Kolmogorov's $4/5$-ths law has not yet been realized but falls short by the asymptotic value by a small logarithmic term, the exponents do not saturate but will approach a constant logarithmically, but the qualitative effect is the same. The precise saturation value also depends on the value chosen for $(2n-1)A_{2n-2,2}/A_{2n,0}$.

We will draw two conclusions based only on the qualitative property that the saturation occurs. First, from Eq.\ (28), it readily follows that the smallest scale of turbulence applicable to large order moments is given by
\begin{equation}
\eta_\infty = L Re^{-1} = \eta Re^{-1/4},
\end{equation}
which at high Reynolds numbers is much smaller than the Kolmogorov scale $\eta$. This has the important consequence that, as the Reynolds number increases, there is an increasing need to resolve DNS better and better---in the limit, up to $\eta_\infty$, as was pointed out many years ago by Yakhot and Sreenivasan \cite{yakh10}.

The second important conclusion is that the characteristic velocity fluctuation for very large $n$ is given by
\begin{equation}
[\langle (\delta_r u)^{2n} \rangle/u_0^{2n}] \propto (r/L)^{\zeta_{2n}/2n}.  
\end{equation}
When $\zeta_{2n}$ tends to a constant with increasing $n$, the right-hand-side is unity, so it is clear that the velocity fluctuation over even the smallest scale can be of the order of the large scale velocity $u_0$. This is an important practical consequence of anomalous exponents.

The result (49) has an implication for the  singularities of the Navier-Stokes equation. In an interesting paper \cite{ches}, the authors demonstrate that the weak solutions of the Navier-Stokes equations are regular provided $d > 3/2$ in $\Lambda_c \propto Re^{-3/(1+d)}$, where $\Lambda_c$ is a parameter that demarcates the Euler-dominated large and inertial scales from the viscosity-dominated dissipation scales. For Kolmogorov \cite{kolm1}, $d = 3$ and the solutions are clearly regular. The solutions are regular even for $d = 3/2$. According to us, 3/2 is the smallest value that $d$ assumes, which assures the regularity of the weak solutions of the Navier-Stokes equations. (The theory uses constant $\varepsilon$ but it is possible that one may need to account for the intermittency of $\varepsilon$ \cite{sree1}.)

A further comment on the saturation of exponents $\zeta_{2n}$, which resembles the situation for Burgers turbulence \cite{burg} with shocks, as well as passive scalars \cite{krai3,pass}. As already implied, there is always a competition between the effect of nonlinearity that tries to steepen fluctuations and pressure that tends to prevent it by distributing energy from one component to another, annulling the steepening effect of nonlinearity. It appears that for moments of high order, pressure is on the average incapable of countering the steepening effects, thus rendering the exponents of velocity increments to saturate (see Section III). Intermittency is the local (and instantaneous) imbalance between the two effects. Saturation is the sign of extreme intermittency when one of them, namely the pressure effects, has become negligible.

\section{Comparisons with experiments and simulation}

\subsection{Anomalous scaling exponents}

In this section, we make comparisons with a few aspects of the theory with experimental and simulations data. Table I shows the anomalous exponents $\zeta_n$ in the inertial range from the present theory (PT) as well as experiment (EXP) and simulations (DNS). The experimental data are from time series using Taylor's hypothesis (i.e., the time increment $\delta t$ in turbulence being advected by a mean flow $U$ is equivalent to the spatial increment, $r = \delta t \times U$. The numbers are obtained by choosing the inertial range to be the one within which the third-order structure function has an acceptably good linear region, according to the 4/5-ths law. We also list the data from the Extended Self Similarity (ESS) analysis of Benzi et al. \cite{benz}. Below that row, we show scaling of structure functions with actual spatial separation (i.e., no Taylor's hypothesis is used) from simulations on a large $8192^3$ grid \cite{iyer}. Finally, comparisons are made also with the geometric model by She and Leveque \cite{shel}. It is clear that the theory is in excellent agreement with the data. 
\begin{table}[h]
\begin{widetext}
\begin{ruledtabular}
\begin{tabular}{ccccccccccccccccc}
$n$ & $1$ & $2$ & $3$ & $4$ & $5$ & $6$ & $7$ & $8$ &$ 9$ & $10$ & $11$ & $ 12$ & $13$ & $14$ \\
\hline
$PT$ & $0.366$ & $0.7 $ & $1.0$ & $1.27$ & $1.525$ &$1.76$ & $1.97$ & $2.17$  & $2.35$ & $2.52$ & $2.68$ & $2.83$ & $2.97$ & $3.1$\\
$EXP $& $-- $ & $0.7$ & $ 1.0$ & $1.25$ & $1.56$ & $1.8$ & $2.0$ & $2.2 $ & $2.3$ & $2.5$\\
$ESS $ & $--$ & $0.7$ &  $1.0 $ & $1.28$ & $1.53$ & $1.77$ & $2.01$ & $2.23$ \\
$DNS$& $-- $& $0.7$ & $ 1.0$ & $1.32$ & $1.56$ & $1.81$ & $2.02$ & $2.21$ & $2.38$ & $2.52$ & $ 2.66$ & $2.74$ & $$ &$$\\
$ER$& $-- $& $3\times 10^{-3}$ & $ 10^{-2}$ & $4\times 10^{-3}$ & $ 10^{-2}$ & $7\times 10^{-3}$ & $2\times 10^{-2}$ & $10^{-2}$ & $5\times 10^{-2}$ & $3 \times 10^{-2}$ & $ 4\times 10^{-2}$ & $4\times 10^{-2}$ & $$ &$$\\
$SL$  & $0.36$ &$ 0.70$& $1.0$ & $1.28$ & $1.54$& $1.78$  & $2.00$ & $2.21$ & $2.41$ & $2.6$ & $2.8$ & $2.9$ & $3.1$ & $3.25$\\
\end{tabular}
\end{ruledtabular}
\caption{Scaling exponents $\zeta_n$ of velocity structure functions $S_{n}\propto r^{\zeta_{n}}$. The rows in order are: (PT) present theory; (EXP) experiment from \cite{sran}, using Taylor's hypothesis; (ESS) exponents for similar data measured using the Extended Self-Similarity \cite{benz}); (DNS) spatial data from large scale direct numerical simulations on an $8192^3$ grid;  (ER) error bars on both sides of the mean values from DNS in the row above, rounded off to the lowest integer level; (SL) geometric model of Ref.\ \cite{shel}.}
\end{widetext}
\end{table}
We show the comparison graphically (Fig.\ 1) between the theory and various data sources listed in the table. We also add an inset which plots $\zeta_n/n$ versus $1/n$ to show the asymptotic saturation behavior for large $n$. The net effect of saturation is clear even though the approach may be slow. No experiment or simulation has shown this saturation compellingly for longitudinal structure functions, but it is clear that they are in agreement with the theory where the two overlap. It must be noted that transverse structure functions are known to display saturation for $n=O(10)$ (see \cite{iyer}).

The next items to compare are the exponents $\rho_n$ and $d_n$ (table 2). From the computation of the inertial range exponents $\zeta_n$, one can compute the exponents $\rho_n$ and $d_n$, using formulae (30) and (31). In experiment and simulations, these two quantities were determined independent of the $\zeta_n$ from table 1, and therefore the comparison with the theory is to be regarded as having an independent value. Considering the complexity of the measurements, the agreement is indeed excellent.

\begin{figure*}
\includegraphics[height=8.0cm]{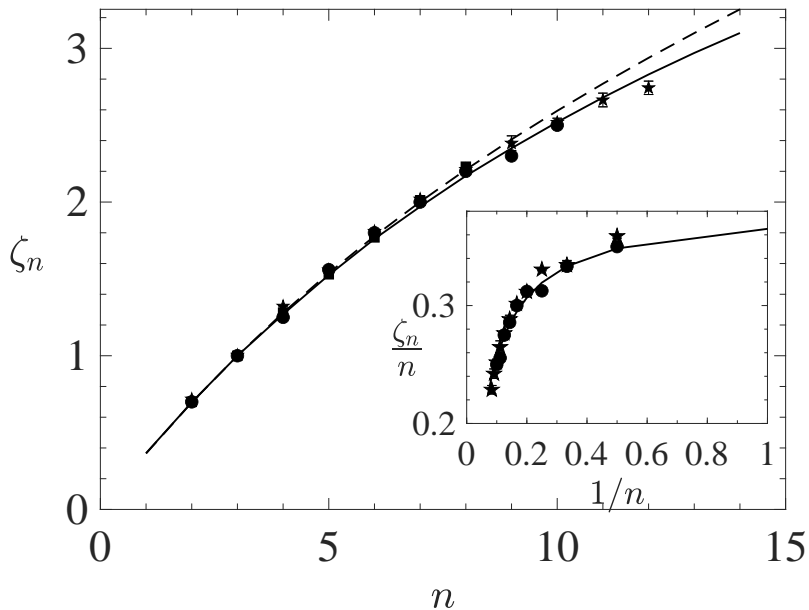}
\caption{Comparison of theoretical predictions (47) for the anomalous exponents $\zeta_n$ with experimental and simulations data. Where the error bars are not visible, they are smaller than the symbol size. The inset shows the ratio $\zeta_n/n$ plotted against $1/n$. It is clear that the saturation tendency is visible in both the theory and experiment as well as simulations}
\end{figure*}

\begin{table}
\begin{center}
\begin{tabular}{ccccccccccc}
\hline
\hline
       & $\rho_n$ & $\rho_{n}$ & $\rho_{n}$ &$d_n$&$d_n$&$d_n$&\\
 $ n $ & Theory & DNS &Experiment& Theory & DNS &Experiment\\
\hline
$ 1$ &  $0.465$& $0.463$& $0.455$ & $0$& $0$& $0$ \\
$ 2$ & $ 1$& $1$&$1$& $ 0.17$& $0.145$&$0.18$\\
$ 3$ & $1.55$& $1.579$ & $1.478$&$0.49$&$0.44$&$0.56 \pm0.06$\\
$ 4$ & $2.17 $& $2.187$ & $2.05$&$0.93$&$0.884$&$0.96\pm0.15$\\
$ 5$ & $2.83$&  $2.817 $& $2.66\pm 0.14$&$1.4$&$1.467$&$$\\
$ 7$ & $ 4.241 $& $4.128$ & $3.99\pm 0.65$\\
\hline
\hline
\end{tabular}
\end{center}
\caption{Comparison of derived exponents $d_{n}$ and $d_n$ with outcomes of DNS and experimental data. Theory: expressions (30)-(31) and (47); DNS data from \cite{donz2}. Experiment from Ref.\ \cite{sran}}.
\end{table}

\subsection{Large scale Gaussian state in fully developed turbulence}

The result on Gaussianity is valid for wavenumbers that are very small compared to the forcing wavenumber, or for length scales that are very large compared to the forcing length. In practice, this seems to work very well for scales of the order of the integral scale. From well-resolved DNS studies of homogeneous and isotropic turbulence presented in \cite{iyer1}, in the largest simulations in boxes of $8192^3$, we compare below the actual moments of velocity differences. The original paper should be consulted for issues on resolution and convergence. The numbers are presented in groups of three as (order of moment 2$n$, DNS moments, theoretical moments for the Gaussian), the last one being simply $(2n-1)$!! The numbers are: (4, 2.98, 3), (6, 14.9, 15), (8, 105.3, 105), (10, 957.2, 945), (12, 10906.5, 10395), and (214, 169208, 135135). Except for $2n$ = 14, amazingly, they are within a few percent of the theoretical values. 

The situation is satisfactory even in a turbulent boundary layer in the fully turbulent part where the effects of outer intermittency are not strong a similar comparison yields the following result \cite{zuba}: (4, 2.83, 3), (6, 14.75, 15), (8, 108.1, 105), (10, 1039, 945) and (12, 13322, 135135). Even though the flow is not strictly homogeneous and isotropic, it is astonishing that large scale of the order $L$ are closely Gaussian. The generality of this result is due to the Central Limit Theorem which is always valid if there are weakly interacting particle or waves, as in an ideal gas \cite{land}. With increase of ``particle concentration", it become unstable and becomes nonlinear and/or intermittent.

Though less extensive, the available evidence in a very high Reynolds number atmospheric boundary layer is also supportive. In Dhruva \cite{dhru1}, for an atmospheric boundary layer at a height of 2 m, corresponding to a microscale Reynolds number of about 5869, the fourth and sixth order structure functions $S_{2n}(r)$ for large $r$ of the order L were found to be 2.9 and 15.2, respectively, very close to $(2n-1)!!$ In an independent but similar measurement at about $30$m above the ground \cite{dhru1}, for which $R_\lambda = 10,340$, the corresponding result for $2n = 6$ was 15. (Higher order data at these high Reynolds numbers are unavailable because of convergence issues.)

We thus conclude that in fully turbulent parts of turbulent shear flows except very close to the wall, the large scale velocity is not far from Gaussian. We believe it is this feature of the large scales in broad range of circumstances that enables low-order engineering models to work well (see, for example, Ref.~\cite{laun1,laun2,bart}). It appears that the large scales are closely Gaussian for conditions (5) and (6) obtained in the weak coupling limit, for which the result is asymptotically exact and can be used as the basis for accurate evaluations of various constants in the inertial range. It also appears to be the case for naturally created shear flows, at least when the Reynolds numbers are very large.

A qualitative but significant point supporting the universality of the Reynolds number of transition of the Gaussian state and the multi-scaling state is the following. Townsend \cite{town} showed that the Reynolds number based on an eddy viscosity is 12.5 for turbulent wakes, and discussed the result in the following physical terms (as interpreted by us). A turbulent flow is essentially on the verge of instability even in the presence of fluctuations which wax and vane in time. At some point in the cycle, the ``mean flow" undergoes a transition at a certain critical Reynolds number, equal to 12.5 for the wakes. (Using the standard relation between the definitions the Reynolds numbers, this number reduces to a microscale Reynolds number of the order 10.) The fluctuations then extract energy from the mean flow, thus increasing the viscosity and reducing the Reynolds number thus stabilizing the flow. The finite amplitude fluctuations in this newly stabilized flow grow with time and the flow becomes unstable again, and the cycle continues. This is true of other flows such as jets as well. Townsend loosely saw this mechanism as the source of large structure in a shear flow. This idea has been developed more explicitly for mixing layers, wakes  and boundary layers in \cite{sree2}. As another anecdotal evidence, we note the demonstration in \cite{aber} that the velocity perturbation in the form of an axial vortex embedded in a boundary layer undergoes inflexional instability at a vortex Reynolds number of 10.  

\subsection{Transition to turbulence}
In the high Reynolds number Gaussian state, suppose we start with a dressed Reynolds number below 9, weak fluctuations develop first; the transition to the anomalous scaling range ensues at a local Reynolds number that exceeds that critical value. This is what we believe happens in the process of transition to turbulence as one increases the Reynolds numbers \cite{schu1,donz1,donz2}. The left panel of Fig.\ 2 shows the predictions are supported best by high-quality DNS in homogeneous and isotropic turbulence. Essentially, these simulations show that the derivative moments $M_n$, which satisfy the Gaussian relation $M_{n}=(2n-1)!! $ for $2 \leq n \leq 6$ for $R_{(\lambda,T)} = Re^* \approx 9$ begin to rise and follow power laws of the type $M_{n}=A_{n}Re^{d_{n}}$, with well-defined exponents $d_{n}$ given by Eq.\ (31). This result is true for turbulence fields driven by a variety of random forces \cite{khur}. The top left figure shows the moments of velocity derivatives $M_{n}=\langle {(\partial_{x}u)^{n}} \rangle/(M_{2})^{n}$. The transition to the anomalous multi-scaling region occurs at $R_{\lambda}\approx 9$. An interesting feature is the Reynolds number dependence of transition with the moments order $2n$ (see right panel of Fig.\ 2). This non-trivial effect is readily explained by the decrease of the volume fraction of the flow occupied by increasingly rarer events \cite{sree1}; see \cite{donz2}. A change in velocity gradient structure observed by \cite{rdas} is also consistent with this picture.

\begin{figure*} 
\includegraphics[height=6cm]{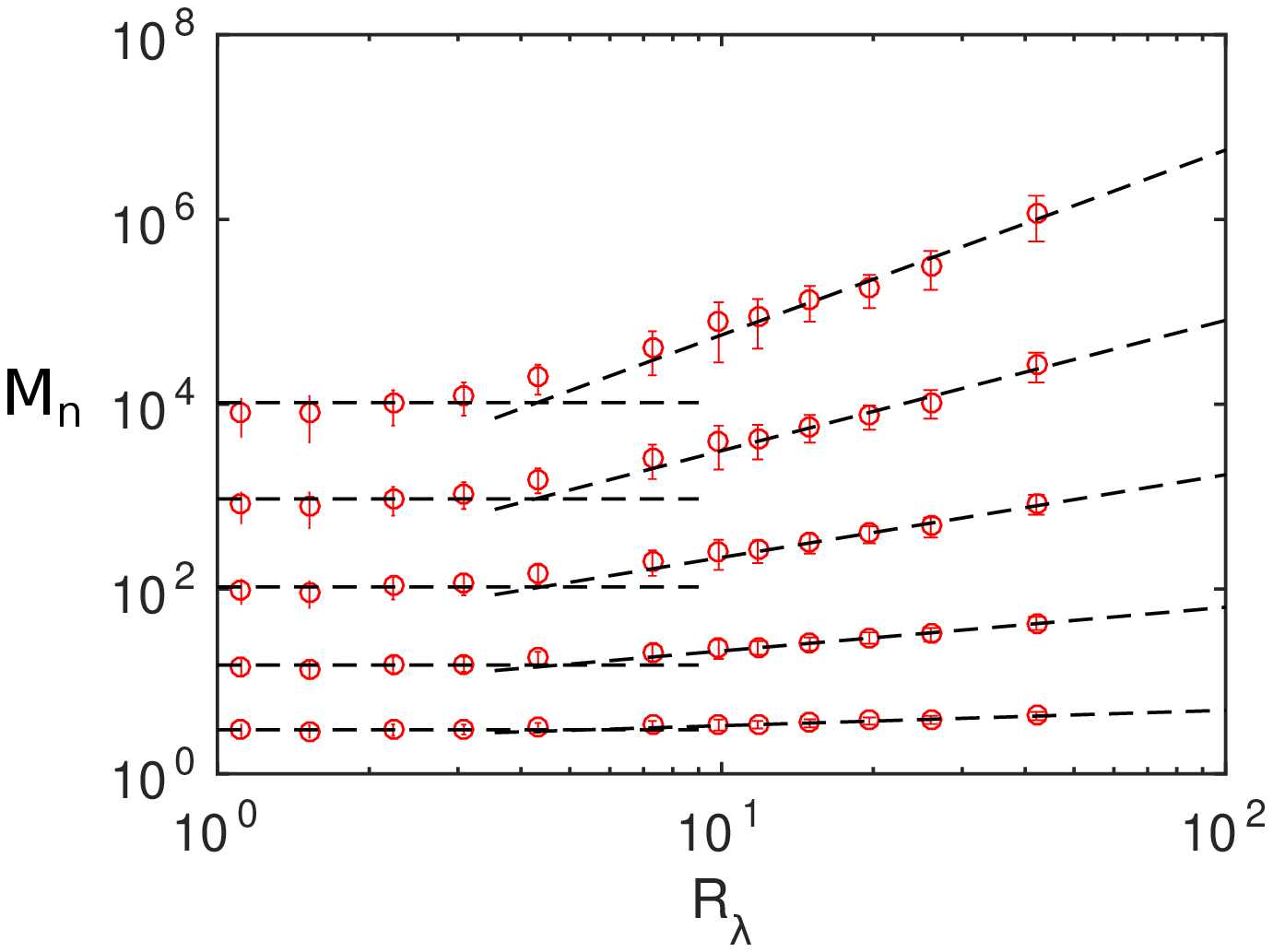}
\includegraphics[height=6cm]{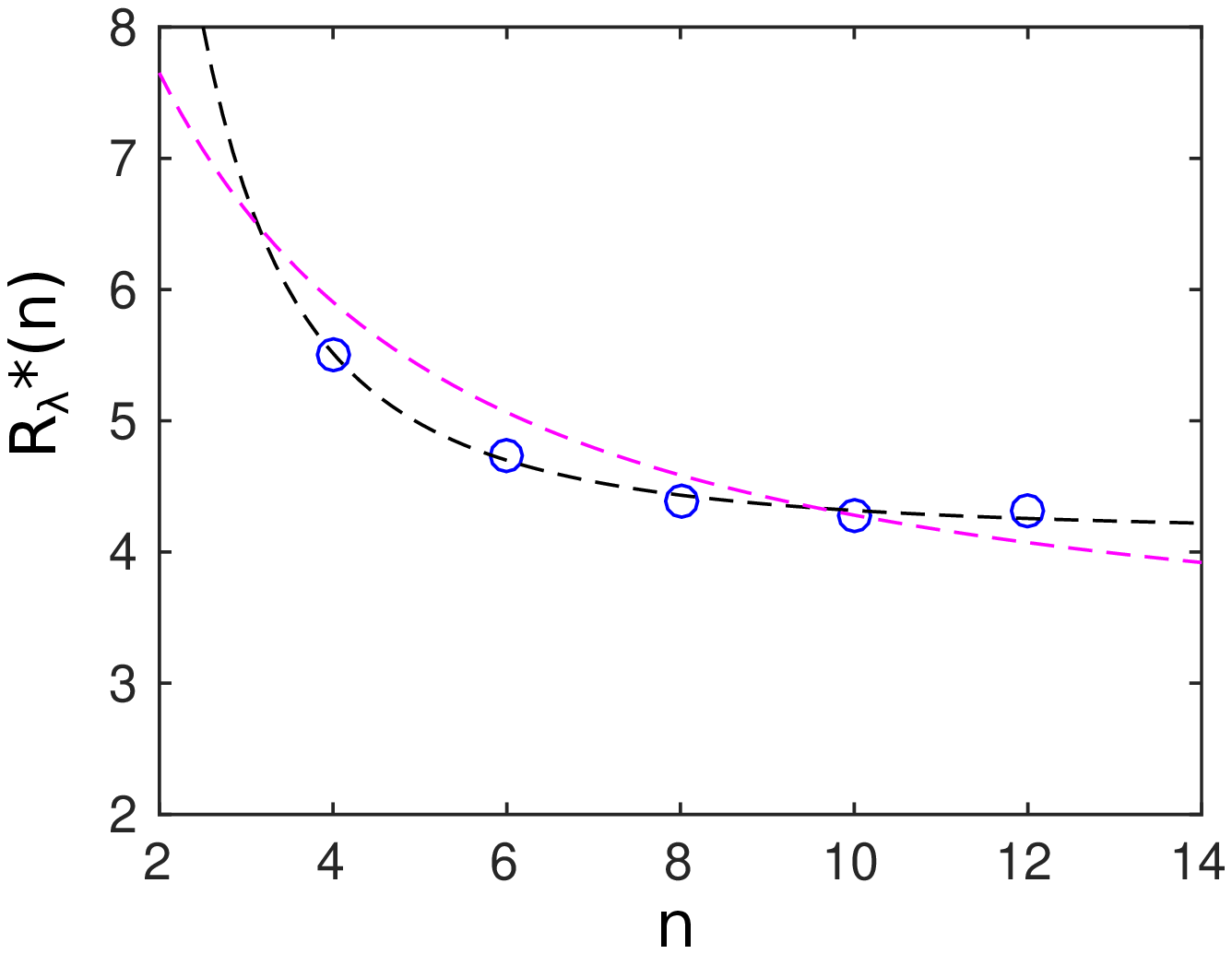}
\caption{Left panel shows normalized moments of velocity gradients $M_{n}$, for $n$ = 2, 3, 4, 5 and 6, of the velocity gradients from direct numerical simulations of the Navier-Stokes equations \cite{donz1,donz2}. For low Reynolds numbers, each of the moments are given by $(2n-1)!!$, i.e., they obey Gaussian statistics, as shown by horizontal lines. Each moment takes off, almost abruptly, to follow the fully developed anomalous scaling relations given by the theory, Eq. (30). The lowest moment undergoes a transition from Gaussian to anomalous state at a Reynolds number of about 9. Higher moments undergo this transition at lower Reynolds numbers $R_\lambda^*(n)$, for which sub-Gaussian fluctuations reflecting a more complex dynamics of forcing, are responsible. These Reynolds numbers are shown on the right panel, where theory (blue dashed line) and simulations (red dashed line) are compared. We do not deal with this particular aspect of the theory in this paper but refer the reader to \cite{donz1,donz2}.}
\end{figure*}

\begin{figure*}
\includegraphics[height=8.0cm]{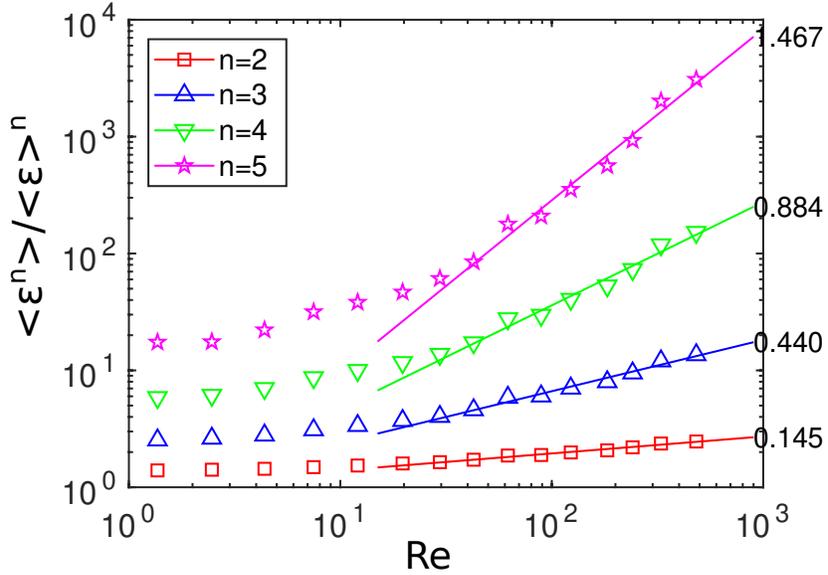}
\caption{Comparison of theoretical predictions for the exponents $d_n$ through the transitional region with numerical simulations of \cite{donz1,donz2}.  The numbers to the right are least square experimental fits to the power-law regions. Table 2 provides comparisons of the exponents.}
\end{figure*}

Finally, as the Reynolds number of the flow increases past the transition points, the exponents $d_n$ can be seen in Fig. 3 to agree with the theoretical result quite well. We have provided only the best fits to the data and not assigned any formal error bars, which would seem unnecessary at this point. 

A further supporting evidence comes from channel flow and thermal convection studies \cite{schu1,schu2,schu3}, though the picture in these two latter flows is somewhat complex. In \cite{schu2}, the authors concluded as follows: ``For small Reynolds numbers ... a transition occurs from sub-Gaussian or nearly Gaussian velocity gradient statistics to intermittent non-Gaussian ones. At the transition Reynolds number the derivative fluctuations are Gaussian." This is the Gaussian state to which we have made a reference. We note that this conclusion is supported by more recent simulations \cite{schu3}. 

An essentially Gaussian state at large scales has been observed frequently in the past for a broader class of conditions. That evidence is worth recalling though it should be regarded as suggestive. Flows lose their stability in successive stages, develop structures, before eventually succumbing to transition. However, the transition to turbulence is not a consequence of successive loss of stability but an abrupt phenomenon, qualitatively resembling the abrupt appearance of the onset of temporal chaos \cite{ruel}. Our theory is oblivious to the generation of well-organized transitional structures, however interesting and important they are in their own right. We recall Busse's \cite{buss} remark that at moderate Prandtl numbers, turbulent convective large scale structures fluctuate in both space and time, despite their well-organized orientation. He remarked that this state combines random processes with the permanence of a large-scale organizing structure. In other words, the initial Gaussian state encompasses all possible structures, and the transition takes place when $R_{\lambda}$ exceeds about 9, this being the last stage of breakdown. In the specific example of the boundary layer, we will not be concerned with the generation of well organized structures such as the Tollmien-Schlichting waves, or the development of three-dimensional structure, the formation of axial vortices, etc., but with the eventual breakdown of the flow; even the precise form of this breakdown does not appear explicitly in our considerations. To give an example, in the boundary layer the scenario one might surmise is the breakdown of the flow to form turbulent spots, as described by Narasimha \cite{nara}; see also Ref.\ \cite{aber}. Our stipulations are that (1) the state prior to the breakdown is essentially Gaussian, and that (2) the breakdown itself happens independent of the formation of instability structures abruptly at an $R_{\lambda}$ of about 9.

\section{Discussion and Conclusions}

It this paper we developed a novel approach for closing the Hopf equations leading to the determination of anomalous exponents in three-dimensional turbulence. For almost 50 years this problem had eluded researchers in the field and was relegated to the list of ``unsolved problems". Very close agreement of the calculated multi-scaling exponents of structure functions and moments of derivatives with experimental data justifies the details of the theory. We note that no uncontrolled model has been used and trust that this effort will contribute to discussion of whether, at long last, we have the essential elements in place for the elusive theory of turbulence for the specific aspects considered here.
 
To reiterate, we considered the Navier-Stokes equations for an infinite fluid driven at a fixed length-scale $\Lambda$, and showed that two physically different intervals: the ``infra-red" limit $r \gg \Lambda$ and the ``ultra-violet" limit $r \ll \Lambda$ overlapped at the energy forcing scale $r=\Lambda$ corresponding to universal Reynolds number, $R_{\lambda} = R_\Lambda \approx 9$. We further showed that the ranges $r\gg \Lambda$ and $r\ll \Lambda$ correspond to ``equilibrium'' and inertial ranges of fully developed turbulence, respectively. Therefore, the forcing scale $\Lambda$ stands for the integral scale of turbulence and, simultaneously, for the mean free path of an equilibrium system. Depending on the Reynolds number $R_{\lambda}$, the flow can be in the either regime. This Reynolds number marks the transition point from the Gaussian to the multi-scaling behavior of strong (ultraviolet) turbulence.

In the inertial range, where viscous effects are zero, the effective or turbulent viscosity $\nu_{T}$, characterizing the energy flux from large-to-small scales, is length-scale independent, equal to $R_{\lambda,T} \approx 8.8$ = constant, independent of both the``bare" viscosity and the forcing power. Since, asymptotically, the Gaussian infrared range can be calculated to all orders in a perturbation expansion, the continuous transition at $R^{*}_{\lambda}\approx 9.0$ means that the entire ultraviolet interval and inertial range are accurately represented by the expressions (8). We have presented available evidence in both the infrared and the ultraviolet region to support the theoretical results.

We wish to stress a few essential points.

1. No experimentally adjustable parameters were involved in the calculations. Representation of dissipation processes in terms of point splitting, connecting dissipation and inertial range dynamics, based on the natural assumption $\partial_{x}u\approx \frac{\delta_{\eta}u}{\eta}$ of the analyticity of velocity field in the dissipation range, is a well-accepted and understood procedure. Also, at the transition point, the universal and scale-independent Reynolds number $R_{\lambda} \approx 9$ provided an additional numerical input used in our calculations.
 
2. Secondly, we had to deal explicitly with the correlation of the pressure gradient and velocity increment in the inertial range  scale $r$. According to Landau and Lifshitz \cite{land} the effective viscosity on this scale is $\nu_{T}\approx r(u(x+r)-u(x)=r\delta_{r}u$. This natural expression appears in the Navier-Stokes equations as a result of scale-elimination from the interval $[\eta, r]$. As a result, in the coarse-grained NS equation, the spatial coordinate $\hat{x}=\delta_{r}x$ enabled us to evaluate $\delta_{r}\partial_{x}p$ and close Hopf equations. The matching conditions on dissipation and integral scales provided the relations needed for closure of Hopf equations. This led to explicit expressions for structure functions in the inertial range and moments of derivatives related to dissipation scales.
 
3. Surprisingly, neglecting time-derivative in the equation the pressure gradient resulted in expression for the ``normal" Kolmogorov scaling $\zeta_{2n}=2n/3$. However, accounting for the contribution of the time derivative turned out to be crucial: the anomalous exponents $\zeta_{n}$ coming from the derived equations, agreed with experimental and numerical data in the entire range $-1<n$. (For negative $n$, we take absolute values of the velocity increments.) Again, no adjustable parameter was involved in the calculation; the only coefficients one needs are $a$ and $b$, but $a=1/2$ from the Navier-Stokes equation for pressure gradient and $b$ is obtained from the theoretically valid constraint $S_3=1$.

4. We found that in the limit $n\rightarrow \infty$ the exponents $\zeta_{2n}$ saturate, resembling the situation in the compressible case of Burgers turbulence as well as passive scalars, both cases containing no pressure. The saturation suggests that the pressure effects are essentially absent for high-order structure functions.
 
5. Finally, we argued that this transition from the Gaussian to multiscaling state is no different from the laminar-turbulent transition and supported it by invoking a number of simulations. This result has now been demonstrated very convincingly for homogeneous and isotropic turbulence for different types of forcing \cite{khur} and for two types of flows \cite{donz1,donz2,schu1,schu2}, in which the Gaussian state transitions to the multi-scaling at $R^*_{\lambda}\approx 9$ as well.  

All these results bring us to ask ourselves if we have the right ingredients of the theory for turbulence in the context of large-scale behavior as well as the scaling exponents. The all-encompassing theory of turbulence may be elusive because some properties of turbulence, such as the behavior of the largest scales of the flow, depends on specifics of initial and boundary conditions, but the essential ingredients of the theory for scales participating in the energetic dynamics appear to be on hand. We have also argued that some specific results are applicable to a variety of other flows as well---so we ask (with humility) whether we are we approaching the solution to the anomalous scaling problem in turbulence theory.\\\\
 
Acknowledgments: This  paper is the result of some 25 years of collaboration between the two authors. Over this time, we have discussed our work with many colleagues too numerous to list here, and are grateful to all of them. We should particularly acknowledge the early influence of the late Steve Orszag, and numerous discussions over the years with H. Chen, D.A. Donzis, G.L. Eyink, K. P. Iyer, S. Khurshid, A.M. Polyakov, J. Schumacher, L. Smith and I.Starosel'sky.

{}


\begin{thebibliography}{}

\bibitem{gait}
C.C. Gaither and A.E. Cavazos-Gaither, Gaither's Dictionary of Scientific quotations, Springer, 2013.

\bibitem{eule} 
L. Euler, Principes g\'en\'eraux du mouvement des fluides [The general principles of the movement of fluids], Mem. de l'Acd. d. Sci. d Berlin 11, 274-315 (1757)

\bibitem{navi} 
M. Navier, M\'emoire sur les lois du mouvement des fluides [Memorandum on the laws of motion of fluids], Mem. de l'Acad. d. Sci. 6, 389-416 (1827)

\bibitem{stok} 
G.G. Stokes, On the theories of internal friction of fluids in motion. Trans. Camb. Phil. Soc. 8, 287-305 (1845)

\bibitem{reyn} 
O. Reynolds, On the dynamical theory of incompressible viscous fluids and the determination of the criterion. Phil. Trans. Roy. Soc. Lond. 186, 123-164 (1894)

\bibitem{wyld}
H.W. Wyld, Formulation of theory of turbulence in an incompressible fluid. Ann. Phys. {\bf 14}, 143 (1961)

\bibitem{moni}
A.S. Monin and A.M. Yaglom, {\it Statistical Fluid Mechanics: Mechanics of Turbulence}, volume II, MIT Press (1975)

\bibitem{krai1}
R.H. Kraichnan, The structure of isotropic turbulence at very high Reynolds number, J. Fluid. Mech. {\bf 5}, 497 (1959)

\bibitem{krai2}
R.H. Kraichnan, Lagrangian history closure approximation for turbulence, Phys. Fluids {\bf 8}, 575 (1965)

\bibitem{orsz}
S.A. Orszag and M.D. Kruskal, Phys. Rev. Let. {\bf 16}, 441 (1966)

\bibitem{laun1}
B.E. Launder and D.B. Spalding, Mathematical Models of Turbulence, Academic Press, New York (1972)

\bibitem{laun2}
B.E. Launder and D.B. Spalding, The numerical computation of turbulent flows, Computer Methods in Applied Mechanics and Engineering {\bf 3}, 269 (1974)

\bibitem{yakh1}
V.  Yakhot and S.A. Orszag, Renormalization analysis of turbulence. I. Basic theory. J. Sci. Computing 1, 3-51 (1986)

\bibitem{yakh2}
V. Yakhot, S.A. Orszag, T. Gatski, S. Thangam and C. Speciale, Development of turbulence models for shear flows by a double expansion technique, Phys. Fluids A {\bf 4}, 1510 (1992)

\bibitem{bart}
C. Bartlett, H. Chen, I. Staroselsky, J. Wanderer, V. Yakhot, Lattice Boltzmann two-equation model for turbulence simulations: High-Reynolds number flow past circular cylinder, International Journal of Heat and Fluid Flow, 42, 1-9 (2013) 

\bibitem{yakh3}
V. Yakhot, Reynolds number of transition and self-organized criticality of strong turbulence. Phys. Rev. E 90, 043019 (2014)

\bibitem{sree1}
K.R. Sreenivasan and C. Meneveau, Singularities of the equations of fluid motion, Phys. Rev. A 38, 6287 (1988)

\bibitem{domb} C. Domb and M.S. Green, Phase Transitions and Critical Phenomena, 1976.

\bibitem{refe}
M. Creutz, P. Mitra and K.J.M. Moriarty, Computer Investigations of the three-dimensional Ising model, J. Stat. Phys. 42, 823-832 (1986)

\bibitem{onsa}
L. Onsager, The distribution of energy in turbulence, Phys. Rev. 68, 286 (1945)

\bibitem{kolm1}
A.N. Kolmogorov, The local structure of turbulence in incompressible viscous fluid for very large Reynolds number, Dokl. Akad. Nauk SSSP 30, 9-13, 1941

\bibitem{poly}
A. Polyakov, A similarity hypothesis in the strong interactions. I. Multiple hadron production in e+e- annihilation, Sov. Phys. JETP {\bf 32}, 296 (1971)

\bibitem{anse}
F. Anselmet, Y. Gagne, E.J. Hopfinger and R.A. Antonia
High-order velocity structure functions in turbulent shear flows,  J. Fluid Mech. 140, 63-89 (1984)

\bibitem{benz}
R. Benzi, S. Ciliberto, C. Baudet, G.R. Chavarria, On the scaling of three dimensional homogeneous and isotropic turbulence. Physica D 80, 385Ð398 (1995)

\bibitem{sran}
K.R. Sreenivasan and R.A. Antonia, The phenomenology of small-scale turbulence. Annu. Rev. Fluid Mech. {\bf 29}, 435 (1997)

\bibitem{dhru}
K.R. Sreenivasan and B. Dhruva, Is There Scaling in High-Reynolds-Number Turbulence? Prog. Theo. Phys. Suppl. 130, 103Ð120 (1998)

\bibitem{chen}
S.Y. Chen, B. Dhruva, S. Kurien, K.R. Sreenivasan and M.A. Taylor, Anomalous scaling of low-order structure functions of turbulent velocity, J. Fluid Mech. 533 183Ð92 (2005)

\bibitem{iyer}
K.P. Iyer, K.R. Sreenivasan and P.K. Yeung, Reynolds number scaling of velocity increments in isotropic turbulence, Phys. Rev. E {\bf 95}, 021101(R) (2017)

\bibitem{iyer1}
K.P. Iyer, K.R. Sreenivasan and P.K. Yeung, Scaling exponents saturate in three-dimensional isotropic turbulence, Phys. Rev. Fluids 5, 054605 (2020)

\bibitem{fors}
D.~Forster, D.~Nelson and  M.J.\ Stephen, Large-distance and long-time properties of a randomly stirred fluid,
Phys.\ Rev.\ A {\bf 16}, 732 (1977)

\bibitem{hopf}
E. Hopf, Statistical hydrodynamics and functional calculus, J. Rat. Mech. Anal. {\bf 1} 87-123 (1952)


\bibitem{poly2}
A.M. Polyakov, Turbulence without pressure,  Phys. Rev. E {\bf 52}, 6183 (1995)

\bibitem{yakh4}
V. Yakhot, Mean-field approximation and a small parameter in turbulence theory. Phys. Rev. E 63, 026307 (2001)

\bibitem{yakh4a}
V. Yakhot, Pressure-velocity correlations and scaling exponents in turbulence. J. Fluid Mech. 495, 135-143 (2003)

\bibitem{yakh5}
V. Yakhot, Probability density and scaling exponents of the moments of longitudinal velocity difference in strong turbulence. Phys. Rev. E 57, 1737 (1998)

\bibitem{kuri}
S. Kurien and K.R. Sreenivasan, Anisotropic scaling contributions to high-order structure functions in high-Reynolds-number turbulence. Phys. Rev. E 62, 2206Ð2212 (2000)

\bibitem{goto}
T. Gotoh and T. Nakano, Role of pressure in turbulence. J. Stat. Phys. {\bf 113}, 855 (2003)

\bibitem{lesl}
V. Yakhot and L. Smith, The renormalization group, the $\epsilon$-expansion and derivation of turbulence models, J. Sci. Comp. {\bf 7}, 35 (1992)

\bibitem{dedo} 
C. De Dominicis and P. C. Martin, Energy spectra of certain randomly-stirred fluids, Phys. Rev. A 19, 419 (1979)

\bibitem{land}
L.D.~Landau and E.M.Lifshitz,  ``Fluid Mechanics'', Pergamon, New York, (1982)

\bibitem{sree84} 
K.R. Sreenivasan, The scaling of the energy dissipation rate. Phys. Fluids 27, 1048 (1984)

\bibitem{kolm2} A.N. Kolmogorov. Energy dissipation in locally isotropic turbulence. Doklady Akad. Nauk SSSR 32, 19Ð21 (1941)

\bibitem{sina}
Y.G. Sinai and V. Yakhot, Limiting probability densities of a passive scalar in a random velocity field. Phys. Rev. Lett. {\bf 63}, 1962-1964 (1989)

\bibitem{yakh6}
V. Yakhot and A. Cheklov, Algebraic tails of probability density functions in the random-force-driven Burgers turbulence. Phys. Rev. Lett. 77,  3118-3121 (1996)

\bibitem{yakh7}
V.Yakhot, Passive scalar advected by a rapidly changing velocity field: Probability density of scalar differences. Phys. Rev. E 55, 329-336 (1997)

\bibitem{yakh8}
V. Yakhot, Reynolds number of transition and self-organized criticality of strong turbulence, Phys. Rev. E 90, 043019 (2014)


\bibitem{kado}
B.B. Kadomtsev, V.I. Petviashvili
On the stability of solitary waves in weakly dispersive media
Soviet Phys Doklady, 15 539-541 (1970)

\bibitem{ches}
A. Cheskidov and R. Shvydkoy, A Unified Approach to Regularity Problems for the 3D Navier-Stokes
and Euler Equations: the Use of Kolmogorov’s Dissipation Range, J. Math. Fluid Mech. 16, 263–273 (2014)

\bibitem{burg}
J. Bec and K. Khanin, Burgers turbulence, Phys. Rep. 447, 1 Ð 66 (2007)

\bibitem{krai3}
R. H. Kraichnan, Anomalous scaling of a randomly advected passive scalar, Phys. Rev. Lett. 72, 1016Ð1019 (1994)

\bibitem{pass}
A. Celani, A. Lanotte, A. Mazzino, and M. Vergassola, Universality and saturation of intermittency in passive scalar turbulence, Phys. Rev. Lett. 84, 2385Ð
2388 (2000)

\bibitem{shel}
Z.-S. She and E. Leveque, Universal scaling laws in fully developed turbulence
Phys. Rev. Lett. 72, 336-339 (1994)

\bibitem{zuba} 
L. Zubair, Studies in Turbulence using Wavelet Transforms for Data Compression and Scale Separation. Ph.D. thesis, Graduate School, Yale University, 1993 (254 pages)

\bibitem{dhru1} 
B. Dhruva, An experimantal study of high Reynolds number turbulence in the atmosphere. Ph.D. thesis, Graduate School, Yale University, 2000 (214 pages)

\bibitem{town} 
A.A. Townsend, The Structure of Turbulent Shear Flow. Cambridge University Press, 1956.

\bibitem{sree2} 
K.R. Sreenivasan, A Unified View of the Origin and Morphology of the Turbulent Boundary Layer Structure, in Turbulence Management and Relaminarisation, eds. H.W. Liepmann and R. Narasimha, Springer-Verlag, pp. 37-61, 1988.

\bibitem{aber}
C.F. Pearson and F.H. Abernathy, Evolution of the flow field associated with a streamwise diffusing vortex. J. Fluid Mech. 146, 2710283 (1984)

\bibitem{yakh10}
V. Yakhot and K.R. Sreenivasan, Anomalous Scaling of Structure Functions and Dynamic Constraints on Turbulence Simulations,
J. Stat. Phys. {\bf 121}, 823Ð841 (2005)

\bibitem{schu1}
J.~Schumacher, K.R.~Sreenivasan and V. Yakhot,
Asymptotic exponents from low-{R}eynolds-number flows, New J.\ of Phys.\  {\bf 9}, 89 (2007).

\bibitem{donz1}
V. Yakhot and D.A. Donzis, Emergence of multiscaling in a random-force stirred fluid, Phys. Rev. Lett. {\bf 119}, 044501 (2017)
 
\bibitem{donz2}
V. Yakhot and D.A. Donzis, Anomalous exponents in strong turbulence, Physica D {\bf 384}, 12-17 (2018)

\bibitem{khur}
S. Khurshid, D.A. Donzis and K.R. Sreenivasan, Emergence of universal scaling in isotropic turbulence (in preparation), 2021

\bibitem{rdas}
R. Das, S.S. Girimaji, On the Reynolds number dependence of velocity-gradient structure and dynamics, J. Fluid Mech. 861, 163-179 (2019)

\bibitem{schu2}
J. Schumacher J,D. Scheel D. Krasnov, D.A. Donzis, V. Yakhot and K.R. Sreenivasan, Small-scale universality in fluid turbulence. Proceedings of the National Academy of Sciences of the United States of America. 111: 10961-5 (200x).

\bibitem{schu3}
J. Schumacher, A. Pandey, V. Yakhot and K.R. Sreenivasan, Transition to turbulence scaling in Rayleigh-B\'{e}nard convection. Phys. Rev. E {\bf 98}, 033120 (2018)

\bibitem{ruel}
D. Ruelle and F. Takens, On the Nature of Turbulence, Commun. math. Phys. 20, 167Ñ192 (1971)

\bibitem{buss}
F.H. Busse, Non-linear properties of thermal  convection, Rep. Prog. Phys. 41, 1929-1967 (1978)

\bibitem{nara}
S. Dhawan and R. Narasimha, Some properties of boundary layer flow during the transition from laminar to turbulent motion, J. Fluid Mech. 3, 418-436 (1958)






\end{thebibliography}
\end{document}